# Title: All-Optical Routing of Single Photons by a One-Atom Switch Controlled by a Single Photon

**Authors:** Itay Shomroni[1†], Serge Rosenblum[1†], Yulia Lovsky[1], Orel Bechler[1], Gabriel Guendelman[1], Barak Dayan[1]*

**Affiliations:**

[1] Department of Chemical Physics, Weizmann Institute of Science, Rehovot 76100, Israel.

*Correspondence to: Barak.Dayan@Weizmann.ac.il

[†]These authors contributed equally.

**Abstract**: The prospect of quantum networks, in which quantum information is carried by single photons in photonic circuits, has long been the driving force behind the effort to achieve all-optical routing of single photons. Here we realize the most basic unit of such a photonic circuit: a single-photon activated switch, capable of routing a photon from any of its two inputs to any of its two outputs. Our device is based on a single $^{87}$Rb atom coupled to a fiber-coupled, chip-based microresonator, and is completely all-optical, requiring no other fields beside the in-fiber single-photon pulses. Nonclassical statistics of the control pulse confirm that a single reflected photon toggles the switch from high reflection (65%) to high transmission (90%), with average of ~1.5 control photons per switching event (~3 including linear losses). The fact that the control and target photons are both in-fiber and practically identical makes this scheme compatible with scalable architectures for quantum information processing.

**One Sentence Summary:** We demonstrate for the first time a dual-input, dual-output, all-optical routing of single photons by single photons.

**Main Text:**
Since the concept of quantized light particles marked the birth of quantum mechanics more than a century ago, photons became a key player in the growing field of quantum information science. The fact that photons normally do not interact with each other has made them ideal for communication of quantum information, yet has prevented so far the realization of deterministic all-optical quantum gates based on single photons. The difficulty to achieve nonlinear behavior at the level of single photons, namely photon-photon interactions, is considered a major challenge also in the realization of quantum networks, in which quantum information processing would be performed by material quantum nodes interconnected by photonic channels *(1–3)*.

Accordingly, considerable effort has been invested in recent decades towards the achievement of photon-photon interactions by the mediation of material systems. This effort was pioneered by the attainment of strong coupling between single atoms and optical microresonators in the context of cavity quantum electrodynamics (cQED) *(4–6)*, in which the tight confinement of light in tiny volumes leads to drastic enhancement of the electric field associated with each photon in the cavity mode. Antibunching has been observed both with atoms *(7,8)* and quantum dots *(9)*, and nonlinear phase shifts of 16° *(10)* and recently even π *(11, 12)* have also been demonstrated with single atoms.

Most notably, based on a scheme proposed by Duan and Kimble *(13)*, which involves microwave manipulation of the atomic ground states, there has been recently a series of works which demonstrated nondestructive measurement of an optical photon *(14)*, a single-photon phase switch *(11)*, and the realization of a quantum gate between flying photons and a single atom, showing atom-photon, photon-photon, and atom-photon-photon entanglement *(15)*, all of which can be directly applied to photonic routing.

Nonlinearities such as electromagnetically-induced transparency *(16)* and Rydberg blockade *(17–19)* have been harnessed for the demonstration of all-optical interactions at lower and lower powers *(20–23)*. In particular, all-optical switching in which a weak gate pulse strongly modifies the transmission of another ('target') pulse has been demonstrated with very few *(24)* and even just one *(25)* photons in the gate pulse. In order to take the next step towards scalable quantum networks, there is a need for multiport all-optical switching schemes that are compatible with photonic circuits.

Here we demonstrate the experimental realization of a robust, simple and scalable scheme for all-optical coherent routing of single photons by single photons, with no need for any additional control fields. Based on a series of theoretical works from the recent decade *(26–33)*, this scheme utilizes the mediation of a single, cavity-enhanced three-level atom in the regime best described as the fast-cavity limit. Like in the well-known regime of strong-coupling, in this regime the coherent coupling rate $g$ between the atom and the cavity mode is larger than both the incoherent loss of the cavity $\kappa_i$ and the atomic spontaneous emission rate $\gamma$. Yet in contrast to strong coupling, in this regime the coupling $\kappa_{ex}$ between the cavity and its input/output modes (the two directions of propagation in the optical fiber, in our case) is the fastest rate in the system. This means, for example, that a photon emitted by the atom will exit the cavity by coupling to the fiber before the atom could re-absorb it. Accordingly, the dynamics in this regime are best described by the tools of open quantum systems *(34)*, and do not include reversible Rabi oscillations between the atom and the cavity mode, but rather irreversible cavity-enhanced spontaneous emission of the atom into the fiber. The crucial parameter in such a system is therefore the Purcell enhancement factor, namely the ratio between the atom's cavity-enhanced emission rate into the fiber, to the free-space spontaneous emission rate $\gamma$. This ratio is typically defined as the cooperativity $C$ and is proportional to the ratio between the cavity's quality factor $Q$ to its mode volume $V$: $C=g^2/\kappa\gamma \propto Q/V$, with $\kappa = \kappa_i + \kappa_{ex}$.

By coupling the atom to high-quality (small $\kappa_i$) micro-resonators with tiny mode volumes, it is therefore possible to reach an enhancement factor $C \gg 1$, in which case the atom can be assumed to be interacting primarily with the fiber, a property that made this regime be coined also as the 'one-dimensional atom' regime *(35)*.

The underlying mechanism in our switching scheme is simple and robust, and is in fact similar to the interference mechanism that makes metallic mirrors reflect light. The free charges in the mirror oscillate in response to the incoming field and radiate both forward and backward a field that is opposite in phase to the incoming one. The result is destructive interference in the forward direction, which leaves the incoming light no other choice than to be reflected backwards *(26)*. The same effect occurs with a two-level atom in the one-dimensional atom regime: the atom radiates in both directions, and the destructive interference with the incoming probe in the forward direction leads to reflection of the probe backwards. The difference between a single atom and a macroscopic mirror is exhibited by the fact that the reflected light in this case is sub-Poissonian, as the atom cannot reflect more than one photon at the same instant *(36)*.

In order to exploit this mechanism for optical switching, one needs to introduce 'memory' into the system, namely a way to make reflection of a single photon toggle the internal state of the atom. This is performed by using a three-level, Λ-type configuration in which the atom has *two* transitions, each coupled to only *one* direction of propagation. Such a configuration is depicted in Fig. 1A, where it is assumed that $\sigma^+$ polarization propagates only to the right, and $\sigma^-$ propagates only to the left.

As shown, assuming the initial state of the atom is at $m_F = -1$ and an incoming $\sigma^+$ probe, the destructive interference between the probe and the $\sigma^+$ field emitted by the atom forces the atom to emit a $\sigma^-$ photon to the opposite direction, thereby deterministically ending at the $m_F = +1$ state. As a result, any subsequent $\sigma^+$ photons will not interact with the atom and accordingly will be transmitted *(32)*. Symmetrically, at this stage the system becomes reflective to $\sigma^-$ photons coming from the right.

We therefore see that this system behaves as a symmetric toggle-switch with two inputs, two outputs, and two internal states (Fig. 2A), in which each reflection toggles the internal state. To implement this system as a photonic switch controlled by single-photon pulses, we simply define the first pulse as the control pulse, and the second one as the target, namely the pulse that needs to be routed from a certain input to a certain output.

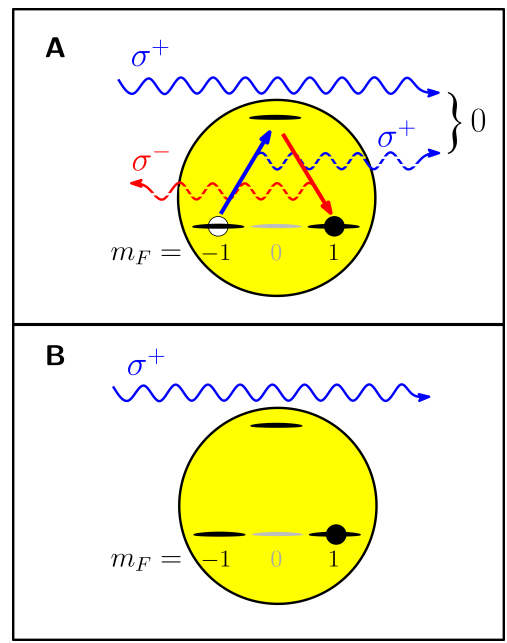

**Fig. 1**: **The two states of the photonic switch.** The $F = 1 \rightarrow F' = 0$ transitions in the D2 line of $^{87}$Rb are used to realize a three-level Λ-configuration. The $\sigma^+$ ($\sigma^-$) transition is coupled only to a photonic mode propagating to the right (left). The atom is irradiated by a $\sigma^+$ probe (solid blue wavy arrow). **(A)** For an atom in the $m_F = -1$ ground state, the incoming $\sigma^+$ photon will be deterministically reflected as a $\sigma^-$ photon due to the destructive interference between the probe and the $\sigma^+$ emission from the atom (dashed blue wavy arrow), thereby toggling the state of the atom to $m_F = +1$. **(B)** An atom in the $m_F = +1$ ground state does not interact with $\sigma^+$ photons, which are accordingly transmitted. The entire scheme is left-right symmetric.

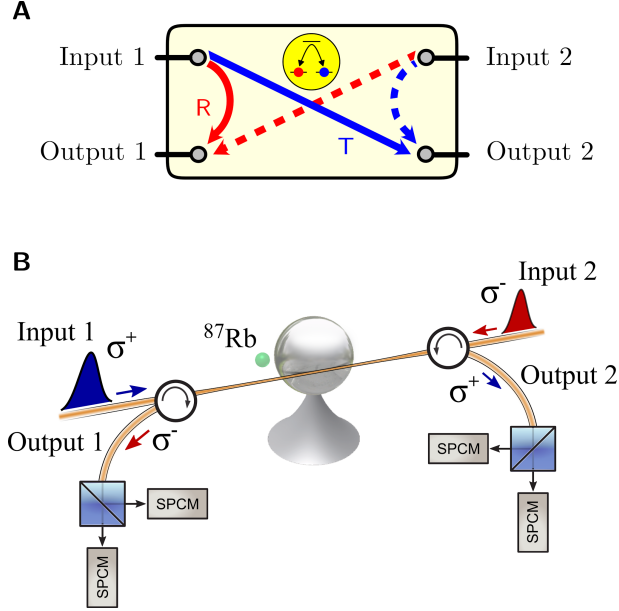

**Fig. 2. Schematic of the photonic switch. (A)** Each input, 1 (solid arrows) or 2 (dashed arrows) may be either transmitted or reflected into one of two outputs, depending upon the state of the atom, at $m_F$ = +1 (blue) or $m_F$ = −1 (red). Each reflection of a single photon toggles the state of the atom. **(B)** The experimental realization of the switch. A single $^{87}$Rb atom interacts with a TM WGM resonance of a chip-based silica microsphere resonator. The input-output interface is provided by a tapered optical nanofiber, with the outputs separated from the inputs by optical circulators *(37)*. $\sigma^+$ modes propagate from left to right, and $\sigma^-$ from right to left. The outputs are delivered to single-photon counting modules (SPCMs).

Figure 2B describes the experimental realization of this scheme. At the heart of our setup is a single $^{87}$Rb atom coupled to a chip-based whispering-gallery mode (WGM) silica microsphere resonator *(37,38)*. Light is evanescently coupled to and from the microsphere by a tapered nanofiber, and photons are detected at the outputs by multiple single-photon counting modules (SPCMs) with temporal resolution of 100 ps. The coupling $\kappa_{ex}$ between the microsphere and the nanofiber is set by careful alignment of their relative position using a piezo positioning system. The overall transmission of the nanofiber and the rest of the optical setup from the microsphere to the SPCMs is slightly above 50%.

Conveniently, by choosing to work with a transverse-magnetic (TM) mode of the microsphere we can approximate to a high degree the situation in which $\sigma^+$ and $\sigma^-$ polarizations are coupled to opposite directions. As recently pointed out by Junge *et al. (39)*, due to a large non-transverse component of the field that has a $\pi/2$ phase shift compared to the transverse one, the polarization in the evanescent wave region is very close to be $\sigma^+$ in one direction, and $\sigma^-$ in the other, with ~96% overlap. The fact that this overlap is not a perfect 100% implies that in this realization our switch has at least 4% chance of reflecting even in the transmitting state (or transmitting during the reflection state), and also that following the reflection of a control photon the atom has at least 4% chance of returning to the initial state, thereby failing to toggle the state *(37)*.

In order to quantify the parameters of our system, we first tuned the WGM resonance and the probe to the cycling transition of the D2 line of $^{87}$Rb ($F = 2 \rightarrow F' = 3$), and adjusted the position of the nanofiber to attain critical coupling with the microsphere, at which, in the absence of an atom, the on-resonance transmission drops to zero (Fig. 3). The observed splitting of the spectrum in the presence of an atom confirms that this resonance is indeed TM *(39)*. From this measurement we deduce a mean coherent coupling rate $g$ = 27 MHz *(37)*.

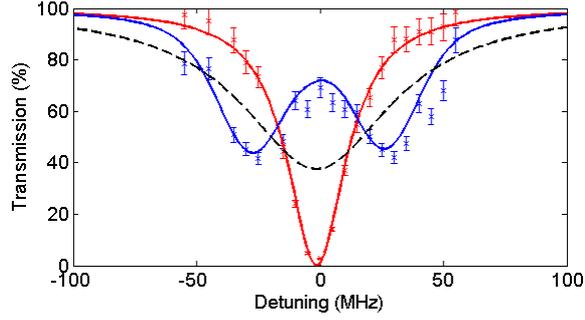

**Fig. 3. Spectrum of the microsphere resonance without (red data points) and with (blue data points) an atom.** In critical coupling, the on-resonance transmission is zero in the absence of an atom. Critical coupling is obtained by setting $\kappa_{ex} = \sqrt{\kappa_i^2 + h^2}$, with $h$ being the coupling between the counter-propagating modes. An atom strongly coupled to the TM mode produces a two-lobed vacuum Rabi splitting *(39)*. The solid curves are theoretical fits giving a mean atom-photon coupling $g = 27$ MHz with a spread of ±10 MHz (owing to variations in the position of the detected atoms), and resonance parameters $\kappa_i = 7.6$ MHz and $h = 1$ MHz. The dashed black curve shows the resonance in the overcoupling regime as used for the photonic switch experiment, with $\kappa_{ex} = 30$ MHz.

For the demonstration of the photon switch we tuned the WGM resonance and the probe to be resonant with the $F = 1 \rightarrow F' = 0$ transition (Fig. 1). As atoms in the state $m_F = 0$ barely interact with the TM mode *(37)*, by choosing this manifold we practically attain the desired Λ-configuration described in Fig. 1, although at the price of decreasing the coherent coupling rate $g$ by a factor of $\sqrt{3}$ compared to the cycling transition.

To approach the fast cavity limit we then increased $\kappa_{ex}$ to 30 MHz. This value was chosen as a compromise between the need to make $\kappa_{ex}$ as large as possible (thus shortening the cavity lifetime and minimizing the linear losses), and the need to maintain large cooperativity C>>1 (to ensure the coherent interaction with the atom is faster than its losses to free space). This value of $\kappa_{ex}$ corresponds to mean cooperativity $C = 2.2$, and to linear losses of 64% in the toroid (i.e. the on-resonance transmission in the nanofiber dropped to 36% of its off-resonance value, see Fig. 3).

The rare event of the presence of one atom within the evanescent wave of the TM mode was identified by sending weak (~2.5 photons per pulse) and short (~15 ns FWHM) pulses in the nanofiber in alternating directions and detecting at least three reflected photons within less than 400 ns *(37)*. Interleaved between the detection pulses were much weaker (0.24 photons on average in each pulse) and longer (~50 ns FWHM) 'target' pulses, whose purpose was to accurately measure the single-photon reflection and transmission properties of the switch. The last detection pulse before each target pulse served as the control, which prepared the atom at a certain initial state. The pulse sequence included control pulses in both directions, thereby preparing the atom half of the times in $m_F = -1$, and half of the times in $m_F = +1$, with the measurement pulse always polarized $\sigma^+$ *(40)*.

The results, composed of ~8000 atom-detection events, are presented in Fig. 4, which shows the measured normalized probabilities for reflection and transmission of the first (and typically the only, if at all) photon in the measurement pulse. These are heralded results, namely they show the statistics of the measurement pulse conditioned on the detection of one reflected photon in the preceding control pulse.

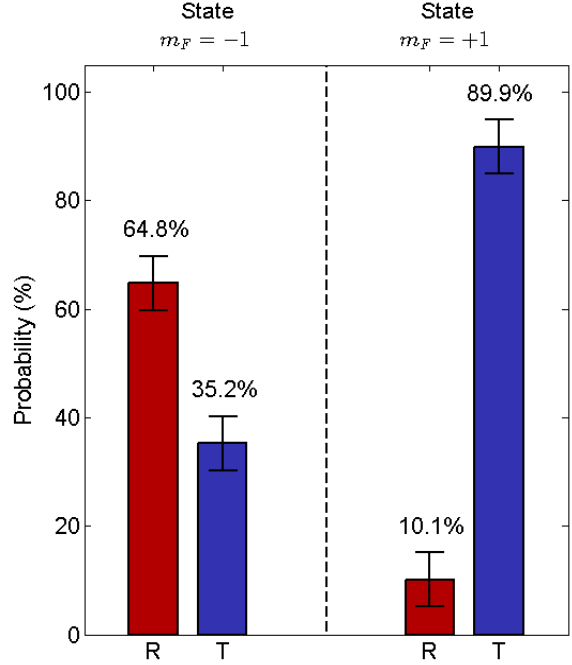

**Fig. 4**: **Single-photon statistics of the photonic switch.** In the $m_F = -1$ state, the probability of a $\sigma^+$ photon to be reflected is 64.8%, dropping to 10.1% in the $m_F = +1$ state.

Specifically, in the cases the control pulse was $\sigma^-$ polarized, detection of a single reflected ($\sigma^+$) photon toggled the switch to a state that reflects $\sigma^+$ photons at probability of 65%. This level is lower than the theoretically expected value of 89% *(37)*. We attribute this difference mostly to the effect of events in which the (untrapped) atom was not present in the cavity mode during the entire measurement. Such events 'mix' into our measurements the statistics of the empty cavity, in which the transmission is nearly 100%, and so they lower the measured probability for reflection considerably even if they occur rarely *(37)*.

In cases the control pulse was $\sigma^+$, detection of a single reflected ($\sigma^-$) photon changed the reflection/transmission ratio by a factor of more than 16, leading to 90% transmission of $\sigma^+$ photons. These are normalized values, namely they reflect the ratio between the photon detection events in both directions. The absolute transmission and reflection values (including measured linear losses in the cavity of 64% at $m_F = +1$, and 51% at $m_F = -1$) are 31.7% reflection and 17.2% transmission in the $m_F = -1$ state, changing to 3.6% reflection and 32.3% transmission in the $m_F = +1$ state. Note that these linear losses are not fundamental, and could be practically eliminated by using higher-$Q$ WGM resonances.

To confirm that indeed the toggling of the state is carried out by a single reflection of a control photon, we look at the statistics of the reflected control pulse. Figure 5A presents the statistics of the second photon (in the cases in which more than one photon was detected) in the control pulse, following the detection of the first reflected photon. As shown, already during the control pulse, following a single reflection the switch becomes highly transmissive, with probability of only 4.4% to reflect a second photon.

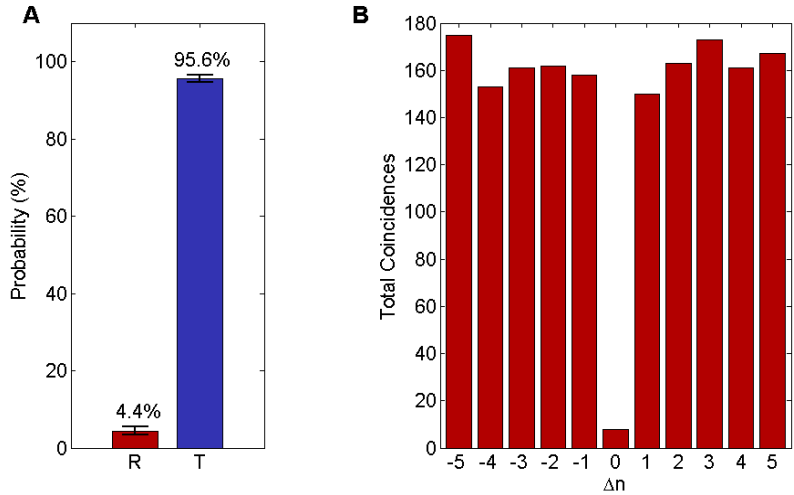

**Figure 5. Single photon nonlinearity within the control pulse. (A)** The statistics of the *second* photon in the control pulse, following the detection of one reflected photon. As shown, the probability of a second reflection drops to 4.4%. **(B)** Antibunching of the reflected control photons, confirming that in most cases only one photon is reflected. For atom-detection events *i* and *j*, the total number of coincident detections of two photons in both *i* and *j* is shown vs. Δ*n* = *i-j*. Within the same pulse the reflection of two photons is accordingly suppressed by a factor of ~20.

This is in fact a drastic demonstration of single-photon nonlinearity, and as such it is highly nonclassical. Accordingly, as presented in Fig. 5B, the reflected control pulses are anti-bunched, meaning that there is a much larger probability for joint detection of two reflected photons (at different SPCMs) in two unrelated control pulses (i.e. from different atom detection events) than within the same control pulse *(37)*.

We can therefore conclude that a single reflection of a control photon toggles the state of the atom with high probability. Taking into account the measured probability for such a reflection (Fig. 4) this corresponds to average of 1.54 control photons per switching event. If we include linear losses by using the absolute reflection probabilities this number increases to ~3.2 photons, yet note that this is an over-estimation, since the control photon can be lost due to linear losses after it already toggled the state of the atom. According to our simulations this occurs at more than a third of the photon loss events *(37)*, leading to average of ~2.5 control photons per switching event in our current realization.

The potential of the scheme presented in this work lies in its compatibility with scalable photonic architectures. The activation of the device is performed only by the single-photon pulses, which are all in-fiber, identical, and routed to the output ports. This means that a routed target photon can serve as the control photon in the next device, or that the same control photon could activate a few devices. Our demonstration therefore paves the way for the application of this scheme as a versatile, robust and simple building-block for a variety of all-optical photonic devices, from quantum memory *(27–29)*, through single photon add/drop filters *(32)*, to photonic quantum gates *(29)*, all of which being completely passive and compatible with scalable quantum networks.

**Acknowledgments:** Support from the Israeli Science Foundation, the Joseph and Celia Reskin Career Development Chair in Physics and the Crown Photonics Center is acknowledged. This research was made possible in part by the historic generosity of the Harold Perlman.

**Supplementary Materials:**

Materials and Methods

Supplementary Text

Figures S1-S2

# Supplementary Materials for

## All-Optical Routing of Single Photons by a One-Atom Switch Controlled by a Single Photon

Itay Shomroni, Serge Rosenblum, Yulia Lovsky, Orel Bechler, Gabriel Guendelman, Barak Dayan

Correspondence to: barak.dayan@weizmann.ac.il

**Materials and Methods**

Microresonator Fabricaton

Our microsphere resonators are fabricated using the same methods for making microtoroidal resonators, first reported in *(38)* and used in other works *(8,36)*. The last stage of fabrication employs a $CO_2$ laser pulse to reflow a silica disk atop a silicon pillar. The microsphere geometry is obtained by starting with a much thinner pillar than that typically used for microtoroids.

Experimental Sequence

At the start of each measurement cycle, we prepare a cloud of ~$30\times10^6$ ultracold atoms in a magneto-optical trap (MOT) and use polarization-gradient cooling to further cool the atoms down to ~7 μK. We then turn off the MOT beams and magnetic field, and let the atoms fall onto the microsphere. About 5 ms before the atoms pass near the microsphere we start the data acquisition stage, which lasts 50 ms (it takes about 15 ms for the entire atomic cloud to fall past the microsphere), during which we send pulse sequences to both inputs of the tapered fiber.

The output channels are separated from the input channels using imbalanced polarizing-beamsplitters that serve as optical circulators. The photons at the outputs are detected using a total of five single-photon counting modules (SPCMs: Perkin-Elmer SPCM-AQRH-14-FC and SPCM-AQ4C-FC). The detection events from the SPCMs are sent to a time-to-digital converter (FAST ComTec MCS6) with a resolution of 100 ps. All events are recorded in computer files for further processing.

Each such measurement cycle takes about one second. The results presented in this work consist of 50,000 such cycles.

## Supplementary Text

Spectrum Measurement

For this measurement, presented in Fig. 3 of the main text, we used a pulse sequence that consisted of on-resonance 'detection' pulses, ~90 ns wide and containing 4 photons on average, interleaved with 'measurement' pulses of variable detunings, ~80 ns wide and containing 1 photon on average. As in the switch experiment, the detection pulses independently ensure the presence of an atom. As the experiment was performed at critical-coupling, atom detection was based on the detection of transmitted photons. We required detection of a total of 4 photons in two successive detection pulses in order to validate the intermediate measurement pulse. The average power in these measurement pulses divided by the transmitted power far from resonance yielded the transmission.

We expect the atom-photon coupling parameter $g$ to vary for different atoms, depending on their distance from the microsphere surface. Accordingly, for the fit to the graph we used a model incorporating a Gaussian distribution of $g$. This yielded a mean value of $g = 27$ MHz with a spread of $\pm 10$ MHz.

The slight asymmetry in the measured data points results from the effect of light-forces on the trajectories of the falling atoms [see *(41)* for more details].

Pulse Sequence and Atom Detection in the photon switch experiment

As described in the main text, our pulse sequence consists of 'detection' and 'target' pulses. The detection pulses verify the presence of an atom by detecting a (single) reflected photon, with the detection pulse preceding the target pulse also serving as the control. The presence of the atom is also verified following the target pulse. Figure S1 shows detection and target pulses for operating the switch in the two states.

Specifically, we use two kinds of pulse sequences within a single measurement, to realize the two states of the switch, as shown in Fig. S1 (A) and (B), respectively. In each case a reflected photon from a control pulse in a different input heralds a different state of the switch. In addition we require the detection of a reflected photon in two pulses following the target, as shown in Fig. S1 by a solid fill. Note that the pulse immediately following the target is not used, as it may depend on the target pulse and thus bias our results. Instead it is used to 'reset' the detection condition (i.e. erase the knowledge on the state of the switch).

False detection of atoms

Using the above described criterion for atom detection yielded false-detection probability of roughly 1.5%, as measured by applying it on data from a time window at which there were no atoms at all. These false detection events lead to the inclusion of the data from measurement pulses in which there was no atom, where the transmission is close to 100%. Accordingly, the measured reflection is lowered by ~1.5%.

However, a much more probable mechanism of false detection, is failure to identify that an atom was present only during the last part of the measurement sequence (during which there are two detection pulses), but not in the first part of the measurement (where there is only one detection pulse, followed by the measurement pulse). By analyzing the distribution of atom-transit durations, we observe that less than half of the atom transits last longer than

300 ns. This effect is a result of the attractive van der Waals forces close to the surface of the microsphere, as described in detail in *(41)*. Accordingly, in half of the cases in which the last two detection pulses identified correctly the presence of the atom, the atom was not present during the first detection/control pulse. As the probability for false-detection is much higher (~24%) if we rely on just one pulse instead of three, the result is that in ~12% of the cases the atom may have 'missed' the first control pulse, and in roughly ~4% of the cases it may have 'missed' the measurement pulse as well. The estimated effect of all these events combined is additional lowering of the measured reflection by ~10%.

Accounting for Detector Afterpulsing

Our single-photon counters (Perkin-Elmer SPCM-AQRH-14-FC and SPCM-AQ4C-FC) have a low probability of generating more than one TTL pulse following photon detection, an effect which is known as afterpulsing. Though most afterpulsing occurs within 50ns following the end of the first TTL pulse, a non-negligible probability of afterpulsing is present during the target pulse, which begins 125 ns after the end of the control pulse (Fig. S1). To account for this effect we characterized afterpulsing in an independent measurement, by sending a single control pulse to each detector. Figure S2 shows detector response with afterpulsing evident as extra photon counts occurring after termination of the pulse. Using this data, we calculated the total probability of a false count during the target pulse to be $5.3 \times 10^{-4}$ and $8.1 \times 10^{-4}$ for the detectors on each output of the photonic switch, respectively. These false counts were then subtracted from the data.

Anti-bunching measurement

Figure 5B displays the cross-correlation of photon detection events from different SPCMs measuring the reflection of the control pulse. There were three SPCMs on this output channel, $D_1, D_2, D_3$, that may detect 0 or 1 photon for atom detection instance $i$. The number of coincident detections in Fig. 5B is calculated by:

$$C(\Delta n) = \sum_i \left[ D_1(i) D_2(i - \Delta n) + D_2(i) D_3(i - \Delta n) + D_1(i) D_3(i - \Delta n) \right]$$

Theoretical Study of the Photonic Switch

As described in the main text, an atom starting in the $m_F = -1$ ground state will reflect an incoming $\sigma^+$ photon (which becomes $\sigma^-$), resulting in the collapse of the atom to $m_F = +1$. This is a result of the destructive interference between the source and the radiation transmitted by atom-cavity system. On the other hand, only the atom is able to reflect a photon, so that no interference will appear in the reflected light. This can be expressed using the input output relations *(32)*:

$$\hat{a}_{out} = \sqrt{2\kappa_s}\,\hat{a}_s + \sqrt{2\kappa_{ex}}\,\hat{a}$$
$$\hat{b}_{out} = \sqrt{2\kappa_{ex}}\,\hat{b}$$

where $\hat{a}_{out}$ ($\hat{b}_{out}$) is the output of the $\sigma^+$ ($\sigma^-$) polarized field, $\hat{a}$ ($\hat{b}$) is the intracavity field of $\sigma^+$ ($\sigma^-$) polarization, and $\kappa_s$ is the inverse width of the input pulse *(32)*. Using the Hamiltonian corresponding to the scheme described in Fig. 1,

$$\hat{H} = -2i\sqrt{\kappa_s \kappa_{ex}} \hat{a}^\dagger \hat{a}_s + \left(g\hat{a}^\dagger \hat{\sigma}_{1e} + g^* \hat{a} \hat{\sigma}_{1e}^\dagger \right) + \left(g\hat{b}^\dagger \hat{\sigma}_{2e} + g^* \hat{b} \hat{\sigma}_{2e}^\dagger \right)$$
$$-i\kappa_s \hat{a}_s^\dagger \hat{a}_s - i\gamma \hat{\sigma}_{ee} - i\left(\kappa_i + \kappa_{ex}\right)\left(\hat{a}^\dagger \hat{a} + \hat{b}^\dagger \hat{b}\right)$$

with $\hat{\sigma}_{ij}$ the atomic dipole operators, we obtain, in the limit of low $\kappa_s$ for the reflection and transmission probabilities

$$P_r = \int_0^\infty \hat{b}_{out}^\dagger \hat{b}_{out} dt = \frac{\kappa_{ex}^2}{\left(\kappa_i + \kappa_{ex}\right)^2} \frac{4C^2}{\left(1+2C\right)^2}$$

$$P_t = \int_0^\infty \hat{a}_{out}^\dagger \hat{a}_{out} dt = \frac{\left(\kappa_i - \kappa_{ex}/\left(1+2C\right)\right)^2}{\left(\kappa_i + \kappa_{ex}\right)^2}$$

where $C = g^2 / \left[\left(\kappa_i + \kappa_{ex}\right)\gamma\right]$ is the cooperativity. Clearly, for $\kappa_i, \gamma \ll g, \kappa_{ex}$ we obtain deterministic reflection, resulting inevitably in the collapse of the atom to $m_F = +1$.

In order to calculate the expected behavior of our system with the experimental parameters ($\kappa_i$, $\kappa_{ex}$, $g$) = (7.6 , 30, (27±10)/√3) MHz, we also have to take into account the nonideal overlap between a forward propagating photon and the $\sigma^+$ mode of the microsphere. Using the analytical solution of the microsphere WGM modes, we estimate the nonideal overlap to result in a parasitic coupling strengths to the $\sigma^-$ and $\pi$ transitions of $\left(g^-, g^\pi\right)$; $\left(g/5, g/7\right)$.

Simulating the resulting dynamics for a Gaussian input pulse, we obtain the following table for the expected behavior of our system, if the atom starts off in $m_F = -1$:

|                       | Reflection | Transmission | Loss  | Total |
|-----------------------|------------|--------------|-------|-------|
| Atomic State Toggle   | 34.3%      | 1.4%         | 16.2% | 51.9% |
| No Atomic State Toggle| 1.9%       | 3.2%         | 43%   | 48.1% |

This corresponds to 89% normalized probability of photon reflection (i.e., given that the photon survived). If the photon is indeed reflected, the atomic state is toggled with 95% probability.

If the atom starts off in $m_F = +1$ the result is

|                       | Reflection | Transmission | Loss  | Total |
|-----------------------|------------|--------------|-------|-------|
| Atomic State Toggle   | 0.1%       | 1.4%         | 0.7%  | 2.2   |
| No Atomic State Toggle| 1.4%       | 33.4%        | 63%   | 97.8  |

This yields 4% normalized probability of photon reflection, and also 4% probability of an unwanted atomic state toggle.

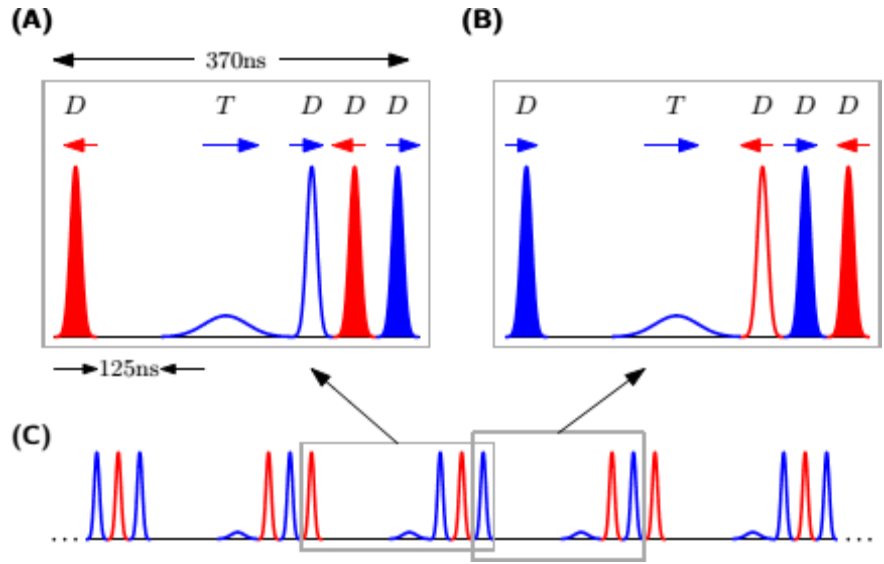

**Fig. S1.** Pulse sequences for operating the photonic switch. Detection pulses *D* are 15 ns with ~2.5 photons on average and target pulse *T* is 50 ns with ~0.24 photons on average. Blue (red) color is used for right- (left-) propagating $\sigma^+$ ($\sigma^-$) pulses. Since the atom changes state upon reflection, successive detection pulses are alternating in direction. **(A)** A left-propagating control pulse sets the switch to reflect the subsequent right-propagating target photon. **(B)** A right-propagating control sets the switch to transmit the target. In (A) and (B) a solid fill indicates the pulses used to detect the atom. **(C)** Zoom out on the entire pulse chain which includes both types of sequences, as indicated by the grey rectangles. The 125 ns wait time following the control pulse is used to reduce probability of afterpulsing.

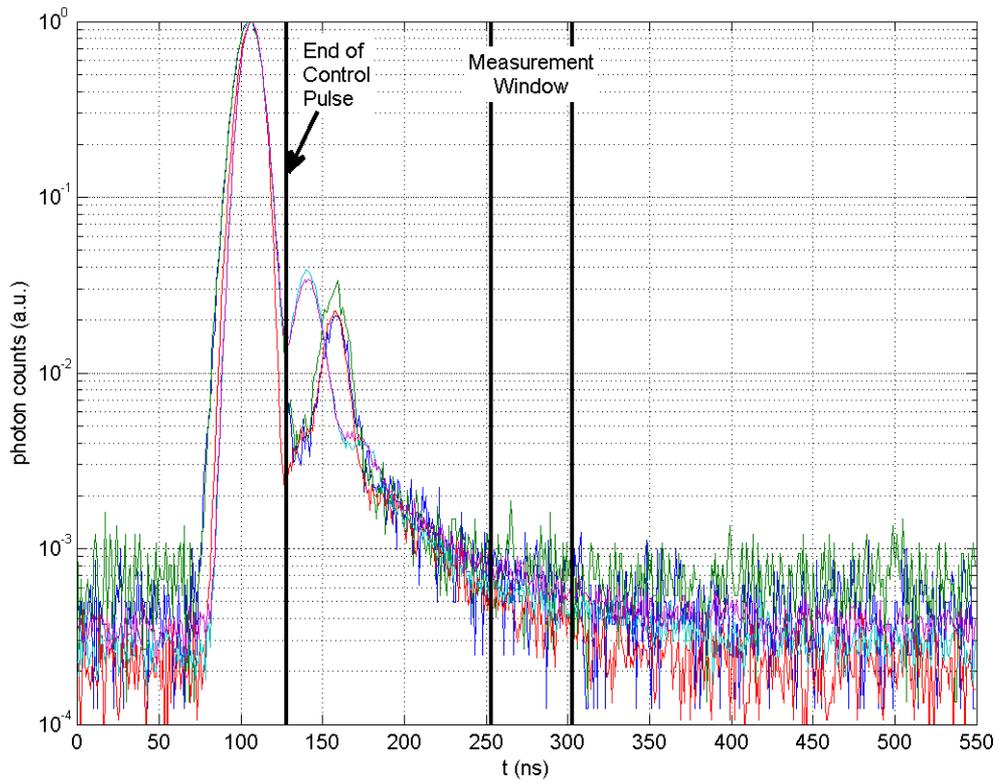

**Fig. S2.** Characterization of SPCM afterpulsing, showing the time frame following a typical control pulse, for all five detectors participating in the measurement. The location of the target pulse, absent in this measurement, is indicated as 'measurement window'. The excessive photon counts following the control pulse are evident.